\documentclass{IOS-Book-Article}

\usepackage{times}
\normalfont
\usepackage{graphicx}
\usepackage{amssymb,amsmath,array}
\usepackage{amsmath}
\usepackage{colordvi}
\usepackage[T1]{fontenc}

\begin{document}
\begin{frontmatter}
\title{Attractor networks and memory replay of phase coded spike patterns}
\author[A,B]{\fnms{Ferdinando} \snm{Giacco}%
\thanks{Corresponding Author: Ferdinando Giacco, Department of Physics, University of Salerno, Via Ponte don
Melillo - I 84084 Fisciano (SA), Italy; E-mail:
giacco@sa.infn.it.}}, and \author[A,B]{\fnms{Silvia}
\snm{Scarpetta} }

\runningauthor{S. Scarpetta et al.}
\address[A]{Dipartimento di Fisica ``E. R. Caianiello'', Università di Salerno, Italy}
\address[B]{ INFN, Sezione di Napoli e Gruppo Coll. di Salerno}
\begin{abstract}
We analyse the storage and retrieval capacity in a recurrent
neural network of spiking integrate and fire neurons. In the model
we distinguish between a learning mode, during which the synaptic
connections change according to a Spike-Timing Dependent
Plasticity (STDP) rule, and a recall mode, in which connections
strengths are no more plastic. Our findings show the ability of
the network to store and recall periodic phase coded patterns a 
small number of neurons has been stimulated. The self sustained dynamics
selectively gives an oscillating spiking activity that matches one
of the stored patterns, depending on the initialization of the
network.
\end{abstract}

\end{frontmatter}
\thispagestyle{empty}
\pagestyle{empty}
\section*{Introduction}
\label{sec_intro}
In many areas of the brain, with different brain functionality, it
has been recently hypothesized that spike phase (i.e.  the
relative phases of the spikes of neurons participating to a
collective oscillation, or the phases of spikes relatively to the
ongoing oscillation) play a crucial role in coding information,
together with the conventional spike rate code. Experimental
evidence of the importance of spike phases in neural coding starts
with the first experiments on theta phase precession in rat's
place cells \cite{Okeefe,Burgess}, showing that both spike rate
and spike phase are correlated with rat's position.  In addition to this, several experiments on short-term memory of multiple objects in
prefrontal cortices of monkeys \cite{MillerPNAS} supported
the hypothesis  that collective oscillations may underlie a
phase-dependent neural coding and that the distinct phase
alignment of information relative to population oscillations may
play a role for disambiguating individual short-term memory items.
hypothesis are also pointed out in the experiments on spike-phase coding of natural stimuli in
auditory and visual primary cortex \cite{panzeri-phase1,panzeri-kayser}.\\
In particular the path-integration system and the hippocampal and
entorhinal cortex circuit, that forms a spatial map of the
environment, has been deeply
investigated \cite{Okeefe,Burgess,GeislerPNAS}, showing that the
place cells and grid cells form a map in which precise phase
relationship among units play a critical role.
The  oscillators interference models
\cite{Blair2008,Burgess-2}  of path-integration are based on the
integration of animal velocity by phase of oscillator cells (such
as a theta cell whose frequency is modulated by the animals'
velocity), and the read-out of this phase by interference between
different oscillators. In the paper of Blair et al.
\cite{Blair2008}, for example, the rate-coded position information
of the grid cells comes from a set of theta oscillatory cells
whose frequency is precisely modulated by the rat's movements'
velocity. Different sets of such theta cells are needed, with
cells in different sets have different frequency, and cells in
each set have a different phase relationships each other.
Moreover, since phase angles between different theta oscillators
encode the rat's position, the oscillators must maintain stable
phase relationships with one other over behaviorally relevant time
scales (many seconds, or dozens of theta cycle periods). Hence,
oscillatory interference models impose strict
constraints upon the dynamical properties of theta oscillators. It
is not presently known whether these constraints are satisfied by
theta-generating circuits in the rat brain, and if so, how.

In this paper we present a possibility to build a circuit in which
stable phase relationships between spikes of different neurons are
maintained in a robust way  with respect to noise. This feature is
due to the robustness of the dynamical attractors with respect to
noise, which are also stable across the changes of frequency.
Indeed the collective frequency of the circuit depends on the firing
threshold $\Theta$  and the phase relationships among
units in the circuit is maintained when output global frequency of
the circuit is changed, indeed it is the phase relationship that
is a dynamical attractor of the circuit and not the absolute spike
timing difference among units.
The mechanism of storing information in the specific spike pattern
of activity and recall info by recalling the specific spike
alignment (or spike phase in case of periodic spatiotemporal
pattern) may be a useful mechanism as substrate for memory.
The importance of precise timing relationships among neurons,
which may carry information to be stored,  is supported also by
the evidence that precise timing of few milliseconds is able to
change the sign of synaptic plasticity. The dependence of synaptic
modification on the precise timing and order of pre- and
postsynaptic spiking has been demonstrated in a variety of neural
circuits of different species. Many experiments show that a
synapse can be potentiated or depressed depending on the relative
timing of the pre- and post-synaptic spikes. This timing
dependence of magnitude and sign of plasticity, observed in
several types of cortical \cite{markram,feldman,Sjostrom} and
hippocampal \cite{Sjostrom,biandpoo,debanne} neurons, is usually termed
Spike Timing Dependent Plasticity or STDP. Here, we face the role of
a learning rule based on STDP in storing
multiple phase-coded memories as attractor states of the neural
dynamics.The spatio-temporal patterns are periodic sequences of spikes,
whose features are encoded in the phase shifts between firing
neurons.\\
We use an Integrate-and-Fire (IF) neuronal model, namely in a Spike-Response
Model (SRM) formulation, which is very popular for theoretical
studies on populations of neurons, especially for large-scale
simulations. This simple choice is suitable to study the storage
and retrieve capability of the network, instead of focusing on the
complexity of the neuronal structure.
Once performed the learning stage, we examine the network
capability to replay(retrieve) a stored pattern. Partial presentation of a
pattern, i.e. short externally induced spike sequences, with
phases similar to the ones of the stored phase pattern, induces
the network to retrieve selectively the stored item, as far as the
number of stored items is not larger then the network storage capacity.
 If the network retrieves one of the stored items, the neural
population spontaneously fires with the specific phase alignments
of that pattern, until external input does not change the state of
the network.
\section{Learning with Spike-Timing Dependent Plasticity}
In the experiment of Markram \cite{markram} it was reported that
if the pre-synaptic spike repeatedly precedes a post- synaptic
action potential within a short time window (10 -20 ms), the
synapse is potentiated (Long Term Potentiation, LTP). If the
opposite occurs, the synapse undergoes depression (Long Term
Depression, LTD). Both effects are combined in a synapse equipped
with STDP \cite{magee, debanne, biandpoo, biandpoo2, markram,
feldman}, where the degree of change in synaptic strength  depends
on the delay between pre and post-synaptic spikes, via a learning
window that is temporally asymmetric (see Fig. 1).
In our model we consider a recurrent neural network with $N(N-1)$
possible connections $J_{ij}$, where $N$ is the number of neural
units. The connections $J_{ij}$, during the learning mode, are
subject to plasticity and change their efficacy according to a
learning rule inspired to the STDP. After the learning stage, the
collective dynamics is studied.\\
According to the learning model already introduced in
\cite{SZJ,NC,PREYoshioka}, the change in the connection $J_{ij}$
that occurs in the time interval $[-T,0]$ due to  periodic spike
trains can be formulated as follows:
\begin{equation}
\delta J_{ij} \propto \int\limits_{-T}^{0}dt
\int\limits_{-T}^{0}dt^\prime \, y_i(t) A(t-t^\prime)
y_j(t^\prime)
\label{lr}
\end{equation}
where $y_j(t)$ is the activity of the pre-synaptic neuron at time
t, and $y_i(t)$ the activity of the post-synaptic one. It means
that the probability that unit $i$ has a spike in the interval
$(t,t+\Delta t)$ is proportional to $y_i(t)\Delta t$ in the limit
$\Delta t\to 0$. The learning window A($\tau=t-t'$) is the measure
of the strength of synaptic change when a time delay $\tau$ occurs
between pre and post-synaptic train. To model the experimental
results of STDP, the  learning window $A(\tau)$ should be an
asymmetric function of $\tau$, mainly positive (LTP) for $\tau>0$
and mainly negative (LTD) for $\tau<0$.\\
While Eqn. (1) holds for activity pattern $y(t)$ which represents
instantaneous firing rate and it has been studied in a analog rate
model \cite{SZJ,NC,PREYoshioka} and in a spin network model
\cite{frontiers2010}, here we want to study the case of spiking
neurons. Therefore, the patterns to be stored are defined as
precise periodic sequence of spikes. Namely, the activity of the
neuron $j$ is a spike train at times $t^\mu_j$,
\begin{equation}
y_j^\mu(t)=\sum_n \delta(t- (t^\mu_j+ n T^\mu)),
\label{spiketr1}
\end{equation}
where $t^\mu_j+n T^\mu$ is the set of  spike times of unit j in
the pattern $\mu$ with period $T^\mu$. Therefore the change in the
connection $J_{ij}$ during the learning of pre-synaptic and
post-synaptic spike trains of the periodic pattern $\mu$, is
given, following Eqn. 1, by
\begin{equation}
J_{ij}^\mu=\sum_{n} A(t^\mu_j-t^\mu_i+ n T^\mu).
\label{lr2}
\end{equation}
\begin{figure}[!htbp]
\centering
\mbox{%
\begin{minipage}[l]{.20\textwidth}
\includegraphics[width=5cm]{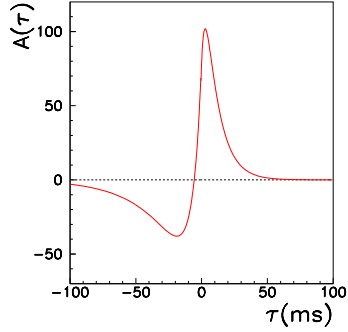}
\end{minipage}%
\qquad \qquad \qquad \qquad \qquad
\begin{minipage}[r]{.50\textwidth}
\caption{a) Plot of the learning window $A(\tau)$ used in the
learning rule (see Eqs. (1), (2), (3)) to model STDP. Parameters
of the function (see Eqn. (4)) fit the experimental data of
\cite{biandpoo}.}
\end{minipage}%
}
\end{figure}
The window $A(\tau)$ that we use, shown in Fig.\ 1, is given by
\begin{equation}
A(\tau) = a_p e^{-\tau/T_p} - a_D e^{-\eta \tau/T_p} \; \textrm{for}\; \tau>0 \; \textrm{and}
\;
A(\tau) = a_p e^{\eta\tau/T_D} - a_D e^{\tau/T_D} \; \textrm{for}\;  \tau<0,
\label{At}
\end{equation}
with the same parameters used in \cite{Abarbanel} to fit the experimental data of \cite{biandpoo},
$a_p = \gamma\,[1/T_p + \eta/T_D]^{-1}$,
$a_D = \gamma\,[\eta/T_p + 1/T_D]^{-1}$,
with $T_p=10.2$ ms, $T_D=28.6$ ms, $\eta=4$, $\gamma=42$.
This function  satisfies the balance condition $\int_{-\infty}^\infty A(\tau) d\tau =0$.\\
Writing Eqn. (1)-(2), implicitly we have assumed that, with periodic
spike trains used to induce plasticity, the effects of all
separate spike pairs  sum linearly with the STDP kernel shown in
Fig. 1. Note that this rule is valid only when, as here, simple
periodic spike trains are used to induce plasticity, and in a
proper range of frequency. Timing-dependent learning curves as
shown in Fig. 1 are indeed typically measured  by giving a
sequence of 100 pairs of spikes repeatedly, with fixed frequency
in a proper range. In fact, pairing single presynaptic and
postsynaptic spikes, or pairing at very low frequency (1Hz) led to
an LTD-only STDP kernel \cite{witt-wang2006}. Similarly, pairing
at high enough frequencies \cite{Sjostrom} the timing-dependent
rule becomes LTP-only, i.e., both positive and negative timings
produce LTP. Moreover the number of pairing also can change the
bidirectional kernel shape into a LTP-only shape. In particular,
for arbitrary non-periodic spike trains major nonlinearities arise
from the history of spike activity, also on timescales longer than
the width of the STDP curve (see \cite{wangshouval2010} and
reference therein). The simple model that we use here, Eqn. 3, is
enough to describe the plasticity that occurs when long periodic
spike trains with frequency in a proper range is used. At very
low, as well as very high frequency, and with few spike pairs, the
timing dependence of plasticity is not well described by the
bidirectional  kernel shown in Fig. 1, and a more detailed model
is needed to account for integration of spike pairs when
not-periodic arbitrary trains are used \cite{wangshouval2010}.\\
\begin{figure}[ht]
\begin{center}
\;\includegraphics[width=4cm]{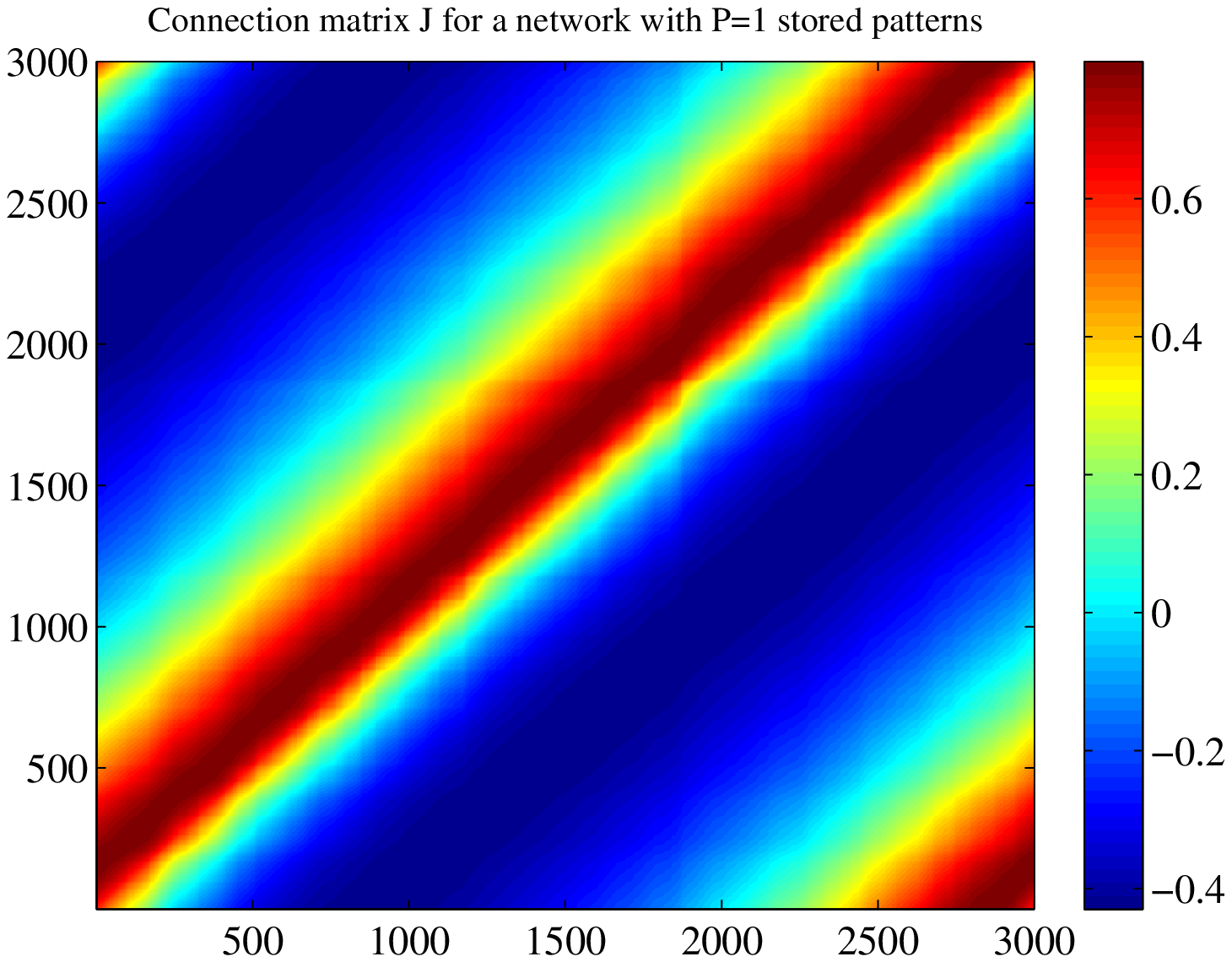}
\;\includegraphics[width=4cm]{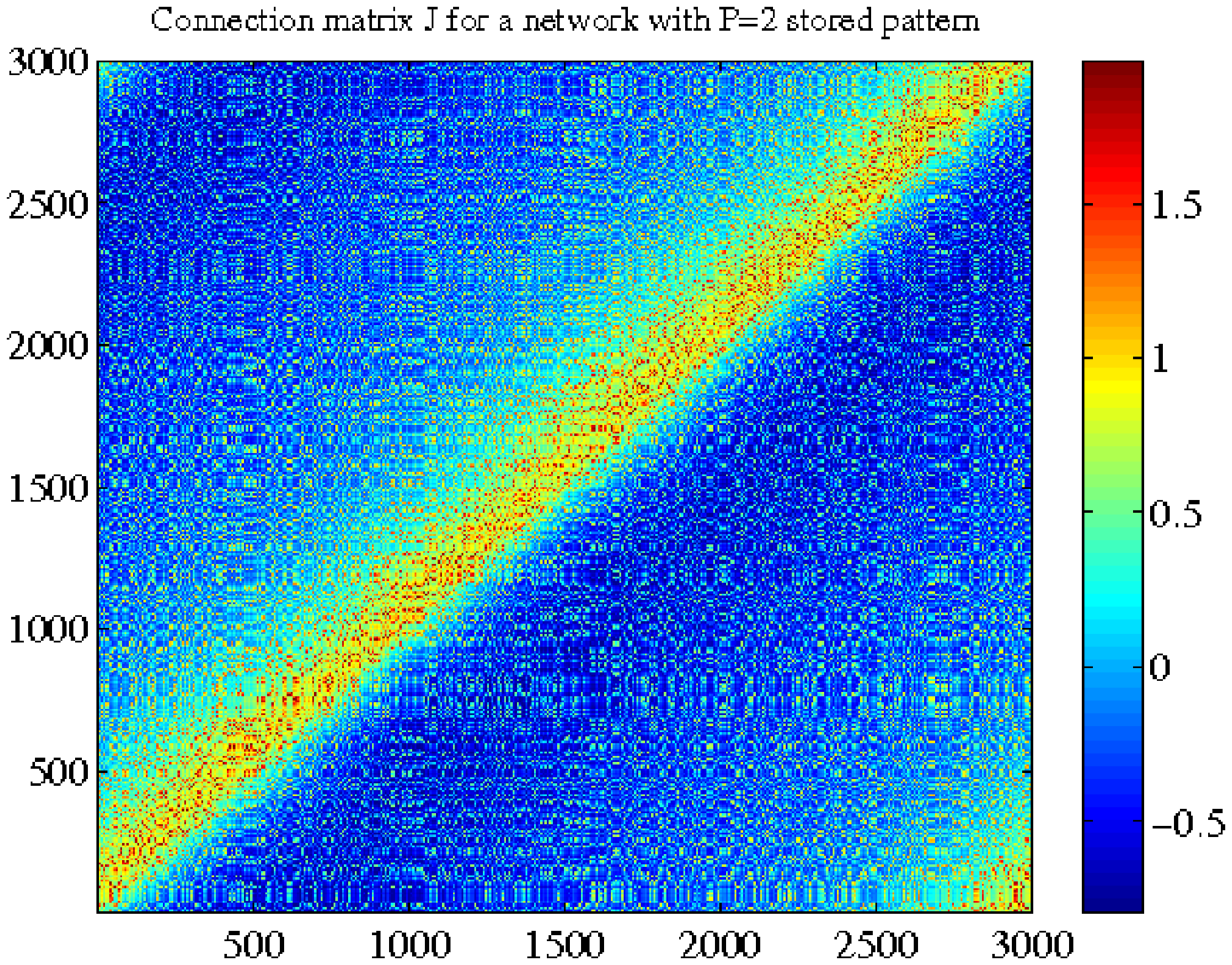}
\;\includegraphics[width=4cm]{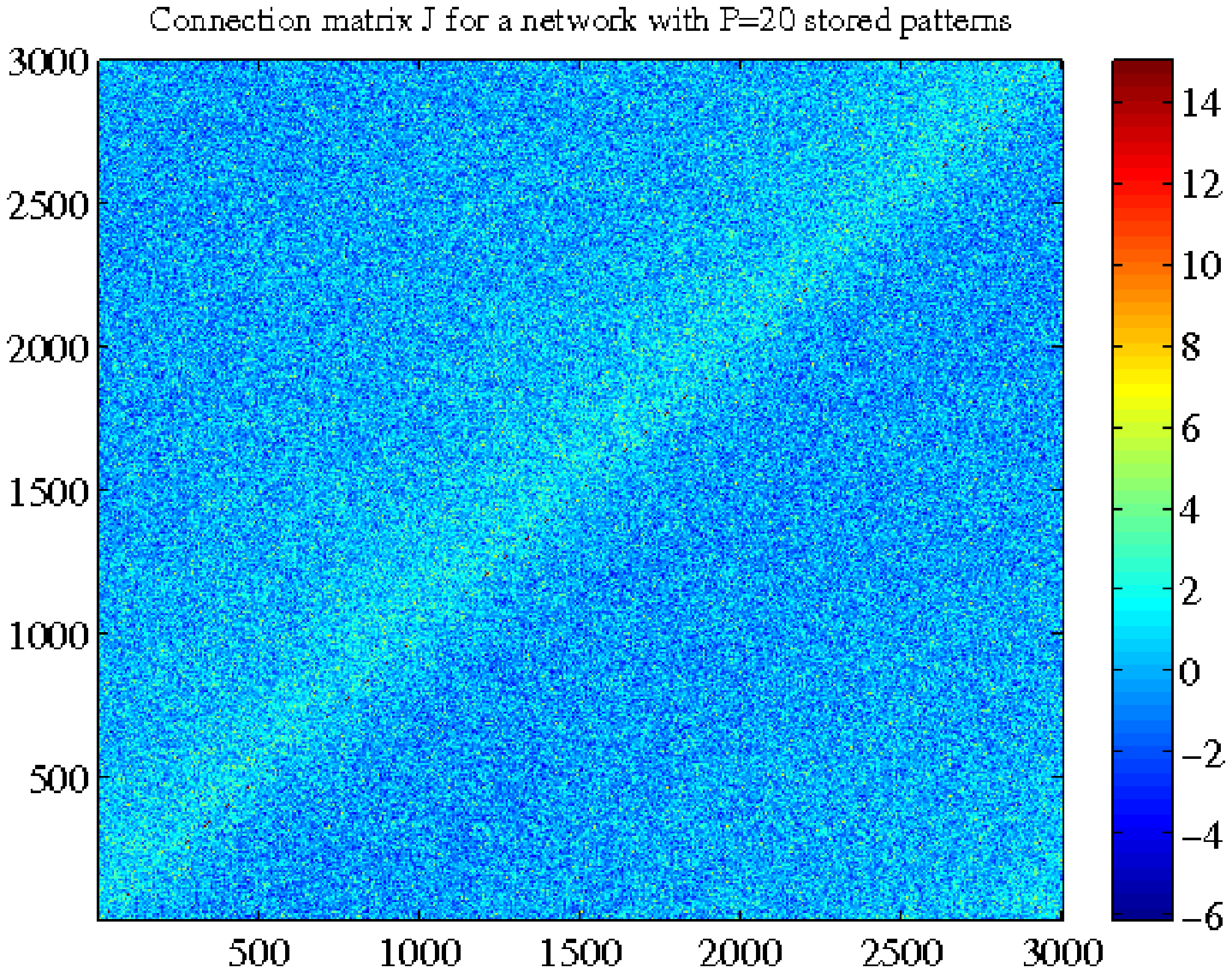}
\end{center}
\caption{ Visual representation of the connection matrix $J_{ij}$
resulting from network with N=3000 units and  P stored patterns at
$\omega_\mu= 20 Hz$. Left P=1, middle P=2, right P=20.}
\end{figure}
The spike spatiotemporal patterns that we study in this paper are
periodic spatiotemporal patterns of spikes, with phases of spike
$\phi^\mu_j$  randomly chosen  from a uniform distribution in
$[0,2\pi)$. Namely, the set of timing of spikes of unit $j$  can
be noted as $ \quad t^\mu_j + n T^\mu =(\phi^\mu_j+2\pi
n)/\omega_\mu $, $\omega_\mu/2\pi$ is the oscillation frequency of
the neurons. Thus, each pattern $\mu$ is defined by its frequency
$\omega_\mu/2\pi$, and by the specific phases of spike
$\phi^\mu_j$ of the neurons $j=1,..,N$.\\
Therefore, the change in the connection $J_{ij}$ provided by the
learning of pattern $\mu$ is given by
\begin{equation}
J_{ij}^\mu=\sum_{n} A(t^\mu_j-t^\mu_i+n T^\mu)=
\sum_n A(\phi^\mu_j/\omega_\mu-\phi^\mu_i/\omega_\mu +2\pi n/\omega_\mu ).
\label{conn}
\end{equation}
In each pattern, information is coded in the precise time delay
between unit $i$ and unit $j$ spikes, that corresponds to a
precise phase relationship among the unit  $i$ and $j$, therefore
this kind of spatiotemporal patterns is often called phase coded
pattern. When we store multiple phase coded patterns defined in
(2), with $\mu=1,2,\ldots,P$, the learned connections are the sum
of the contributions from individual patterns, namely
\begin{equation}
J_{ij}=\sum_{\mu=1}^P J^{\mu}_{ij}.
\label{connP}
\end{equation}
The connection matrix $J_{ij}$ coming out from Eqn. (3) and (5) at
$\omega_\mu=20 Hz$ and $\phi^\mu_i$ randomly chosen in $[0,2\pi]$,
is shown in Fig. 2 for P=1, P=2 and P=20.  The units $i,j$ on the
axes are sorted according to the value of $\phi^1_i$ of first
pattern $\mu=1$. With P=1 it's clearly visible the structure of
the connectivity matrix, however note that even at P=20, when the
correlation structure of the connectivity matrix with the stored
patterns is not visible, the network is still able to selectively
retrieve each of the P stored patterns, in a range of neuronal
threshold values such that storage capacity is equal or higher
then 20.
\section{Model Dynamics}
We distinguish a learning mode in which plasticity rule (3), (4)
and (5) is used to store  P phase-coded pattern into the network
connectivity, from a dynamic mode (or retrieval mode) in which
connections $J_{ij}$ are fixed to the value found after learning
(Eqn. (5)) and the dynamics of the neurons is studied.
Therefore we simulate a Leaky Integrate and Fire network, with
 fixed connections. The Leaky Integrate and Fire model of
single neuron is given by a simple Spike-Response-Model
formulation (SRM) introduced by Gernster in \cite{bookG,SRM}.
While integrate-and-fire models are usually defined in terms of
differential equations, the SRM expresses the membrane potential
at time $t$ as an integral over the past. When membrane potential
reach a threshold $\Theta$ a spike is scheduled. This allows us to use
a event-driven programming and makes the numerical simulations
faster with respect to a differential equation formulation.
In its simplified version \cite{bookG}, the SRM$_0$ model, where neuronal
refractoriness is not taken into account, the internal state of a
spiking neuron  depends on the last output spike and on the total
postsynaptic potential. Supposing the membrane resting potential
is set to zero after a spike, neglecting the shape of the spiking
pulse, the postsynaptic membrane potential is given by:
\begin{equation}
h_i(t)=\sum_{j}J_{ij}\sum_{\hat{t_j}}  \epsilon(t-\hat{t_j}),
\label{IF}
\end{equation}
where the sum over $\hat{t_j}$ runs over all presynaptic firing
times. The function $\epsilon$ describe the response kernel to
incoming spikes on neuron $i$. Namely, each presynaptic spike $j$,
with arrival time $\hat{t_j}$, is supposed to add to the membrane
potential a postsynaptic potential of the form $J_{ij}
\epsilon(t-\hat{t_j})$, where
\begin{equation}
\epsilon(t-\hat{t_j})= K
\left[\exp\left(-\frac{t-\hat{t_j}}{\tau_m}\right) -
\exp\left(-\frac{t-\hat{t_j}}{\tau_s}\right) \right]
H(t-\hat{t_j}) \label{tre}
\end{equation}
where $\tau_m$ is the membrane time constant (here 10 ms),
$\tau_s$ is the synapse time constant (here 5 ms), $H$ is the
Heaviside step function, and K is a multiplicative constant chosen
so that the maximum value of the kernel is 1. The sign of the
synaptic connection $J_{ij}$ set the sign of the postsynaptic
potential's change. When the postsynaptic potential of neuron $i$
reaches the threshold $\Theta$, a postsynaptic spike is
scheduled, and postsynaptic potential is reset to the resting
value zero. Note that a change of $\Theta$ in our model may
correspond to a change in the value of spiking threshold of the
units, or to a global change in the scaling factor of synaptic
connections $J_{ij}$ since what matters is the ratio $J_{ij}/\Theta$.
Anyway a lower value of $\Theta$
correspond to a higher excitability of the network. We simulate
this simple model with $J_{ij}$ taken from the learning rule
given by Eqn. (5)-(6), with $P$ patterns in a network of $N$ units.\\
In the following, the network capacity is analyzed  considering
the maximum number of patterns that the network is able to
perfectly recall. In particular we investigate the role of two
parameters of the model: the frequency of the stored patterns
$\omega_\mu$,  and the firing threshold $\Theta$ which change
the excitability of the network.
\begin{figure}[t!]
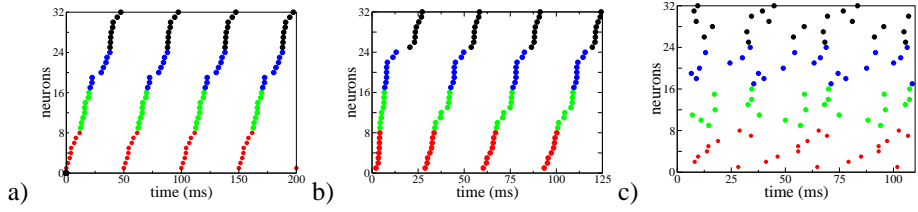

\begin{center}
a)\;
\includegraphics[width=3.5cm]{stored_old20hz.eps}
b)\;
\includegraphics[width=3.5cm]{replayed_old.eps}
c)\;
\includegraphics[width=3.5cm]{notreplay_last.eps}
\end{center}
\caption{ Dynamics of a network with $N=3000$ neurons and
connections given by Eqn. (6) with $P=5$ and $\omega_\mu= 20$ Hz.
A subset of 32 neurons is chosen and sorted by increasing values
of phase $\phi_i^1$ of the stored pattern $\mu=1$. Different
colors refer to phase values belonging to different phase
intervals of pattern $\mu=1$: $(0-\pi/2)$ red, $(\pi/2-\pi)$ green,
$(\pi-3/2\pi)$ blue,$(3/2\pi-2\pi)$ black . The stored
pattern is shown in a), plotting the times $(\phi_i^{1}+2\pi n)/\omega_{1}$. 
Hence, the generated dynamics when a short train of
$M=300$ spikes is induced on the network, corresponding to the pattern 
$\mu=1$ in b) and $\mu=2$ in c). Figure b) shows that when
the network dynamic is stimulated by a partial cue of pattern
$\mu=1$, the neurons oscillate with phase alignments resembling
pattern $\mu=1$, but at different frequency. Otherwise, in  c),
when the partial cue is taken from pattern $\mu=2$, the neurons
phase relationships, even if periodic,  are uncorrelated with
pattern $\mu=1$, and recall the phase of pattern $\mu=2$.} \label{fig_varphi}
\end{figure}
\section{Storage capacity analysis }
We did numerical simulations of the SRM network described in Eqn.
(8)-(9) with $N=3000$ neurons, and connections $J_{ ij}$ given by
(5) with different number of patterns P. After the learning
process, to check if it's possible to recall one of the encoded
patterns, we give an initial signal equal to $M\ll N$ spikes,
taken from the stored pattern $\mu$, and we check that after this
short signal the spontaneous dynamics of the network gives
sustained activity with spikes aligned to the phases $\phi_i^\mu$
of pattern $\mu$. During the retrieval mode, connections strength
is no more plastic as it happens for the learning mode. This
distinction in two stages (learning and retrieval), even though is
not well assessed in real neural dynamics, is useful to simplify
the analysis and also finds some neurophysiological motivations
\cite{Hasselmo,Hasselmo2}.
An example of successful selective retrieval process is shown in
Fig. 3 where, depending on the partial cue presented to the
network, the phase of firing neurons resemble one or another of
the stored patterns. The network dynamic is initially stimulated
by an initial short train of $M=300$ spikes (10\% of the network)
chosen at times $t^\mu_i$ from pattern $\mu=1,2$ and we check if
this initial train triggers the sustained replay of pattern
$\mu=1,2$ at large times.
We introduce a quantitative similarity measure to estimate the overlap
between the network activity during the spontaneous dynamics and the stored phase-coded pattern, defined as
\begin{equation}
|m^\mu(t)|= \left|\frac{1}{N}\sum_{j=1,\ldots,N} e^{-i 2 \pi t_j^*/T^*}  e^{i \phi_j^\mu}\right|
\label{nn}
\end{equation}
where $t_j^*$ is the spike timing of neuron $j$ during the
spontaneous dynamics, and $T^{*}$ is an estimation of the period
of the collective spontaneous dynamics. The overlap in Eqn. (9) is
equal to $1$ when the phase-coded pattern is retrieved perfectly
(even thou with a different time scale), while is $\simeq
1/\sqrt{N}$  when phases of spikes are uncorrelated to the stored
phases.
\begin{figure}[ht]
\begin{center}
a)
\includegraphics[width=3.5cm]{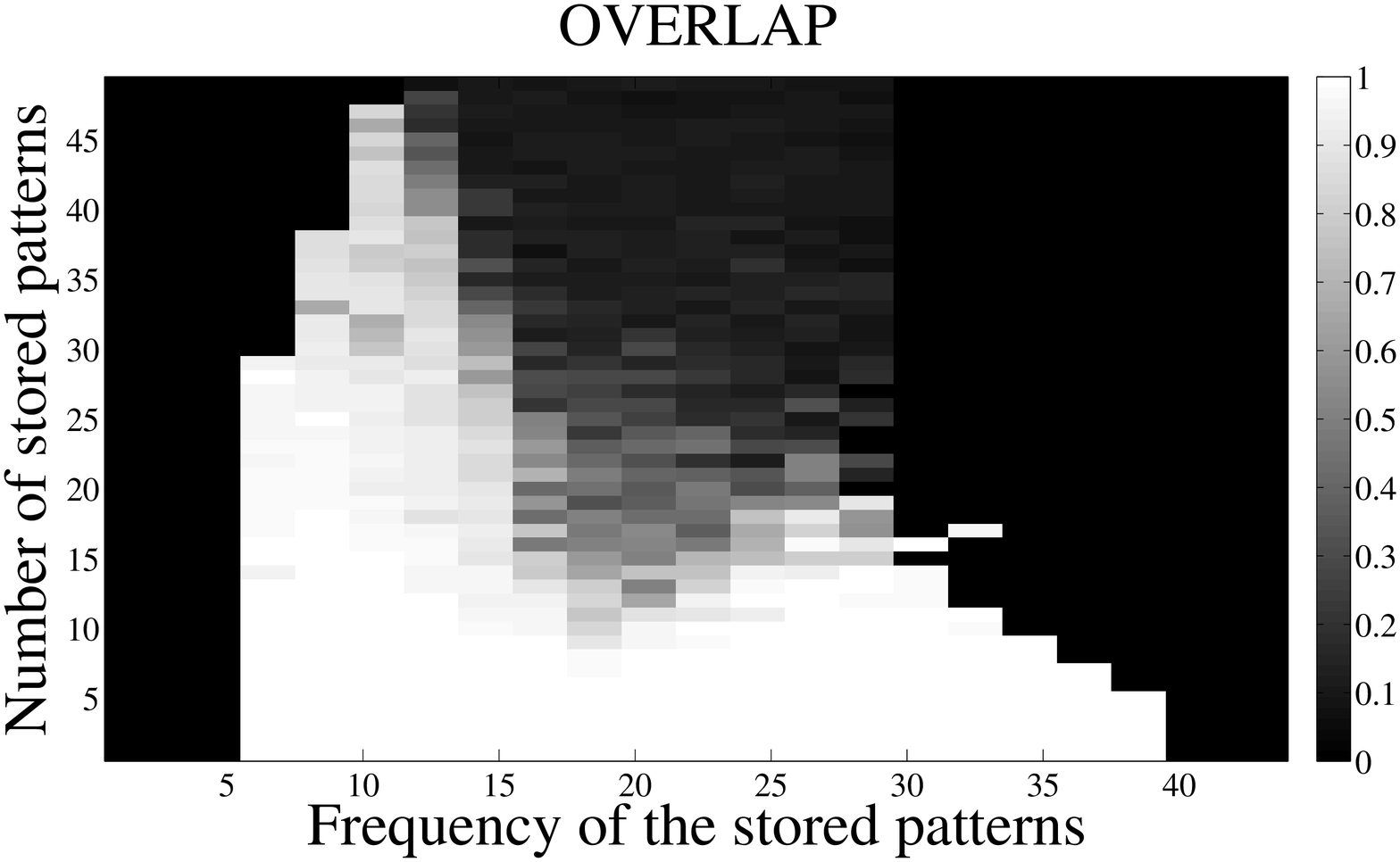}
b)
\includegraphics[width=3.5cm]{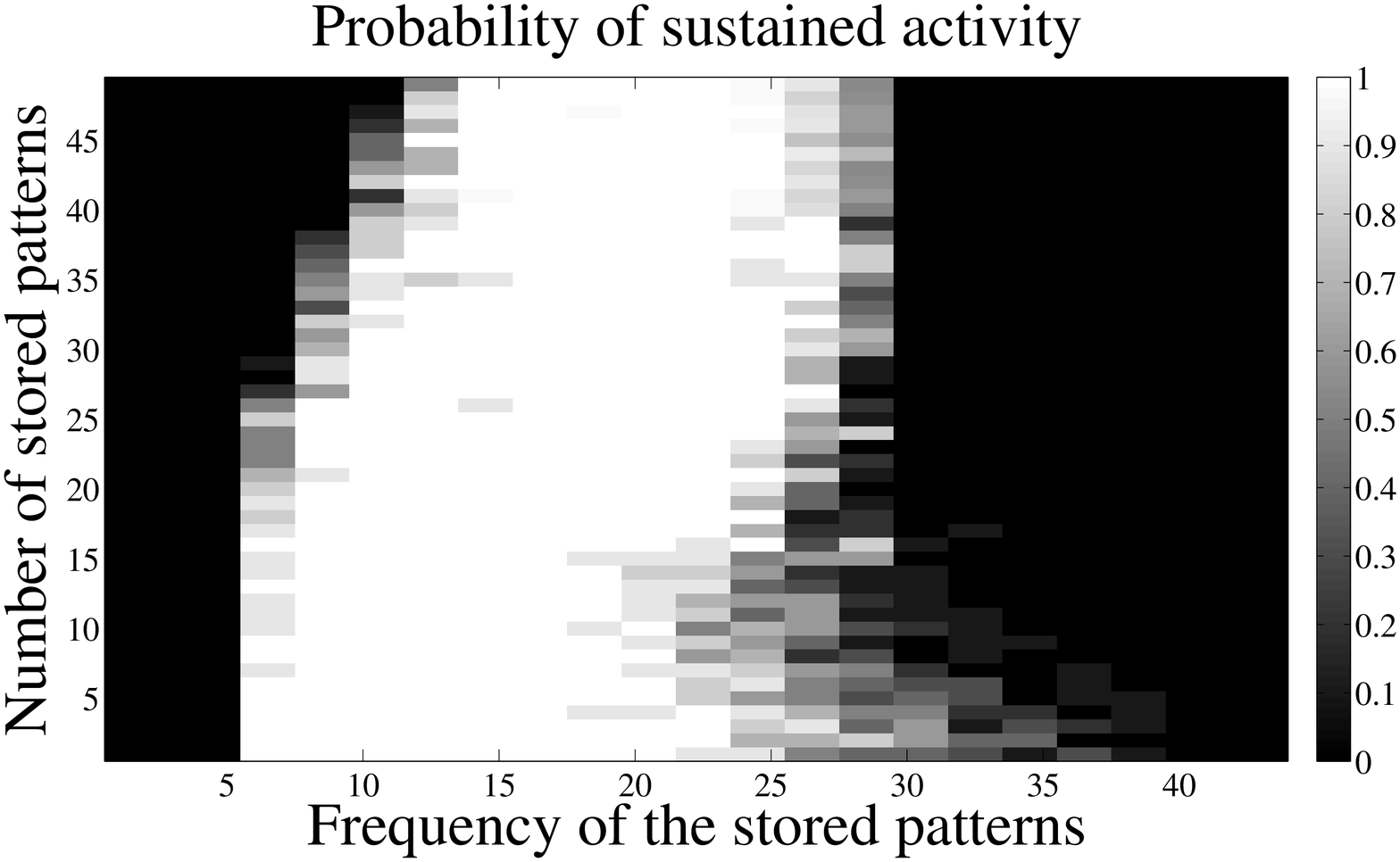}
c)
\includegraphics[width=3.5cm]{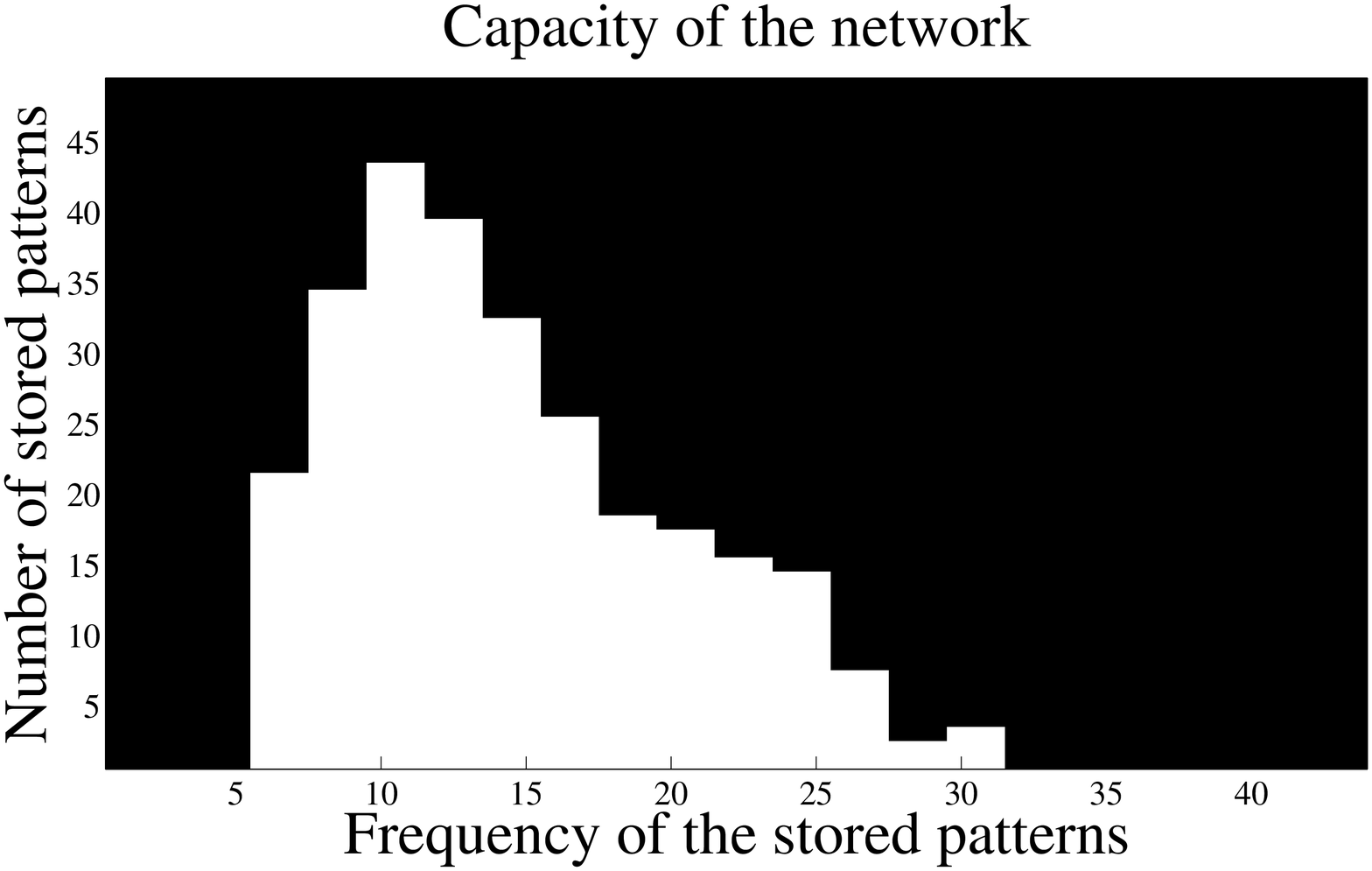}
\end{center}
\caption{ Capacity estimation of a network made up of 3000
neurons. In order to verify that recall of a stored pattern is
correct, we compared the phases of  pattern $\mu=1$ with
spontaneous network dynamic stimulated by 300 spikes chosen at
times $t^{1}_i$ ($i=1\ldots300$). The firing neuronal firing
threshold, $\Theta_{n}=170$. For each boxes in the figure the
y-axis represent the number of stored patterns and x-axis is the
frequency of the stored patterns $\omega_\mu/2\pi$. In a) the
overlap measure defined in Eqn. (9) is shown in a black and white
color legend: the brighter is the color the higher is the overlap
$|m^\mu(t)|$. b) Probability of sustained activity with the same
color legend used for the overlap. In c) the effective capacity of
the network is shown, i.e. a pattern is considered perfectly
retrieved once the product of the overlap with the probability of
sustained activity is higher than 0.5.}
\label{fig_cap}
\end{figure}
\begin{figure}[ht]
\begin{center}
a)
\includegraphics[width=4.5cm]{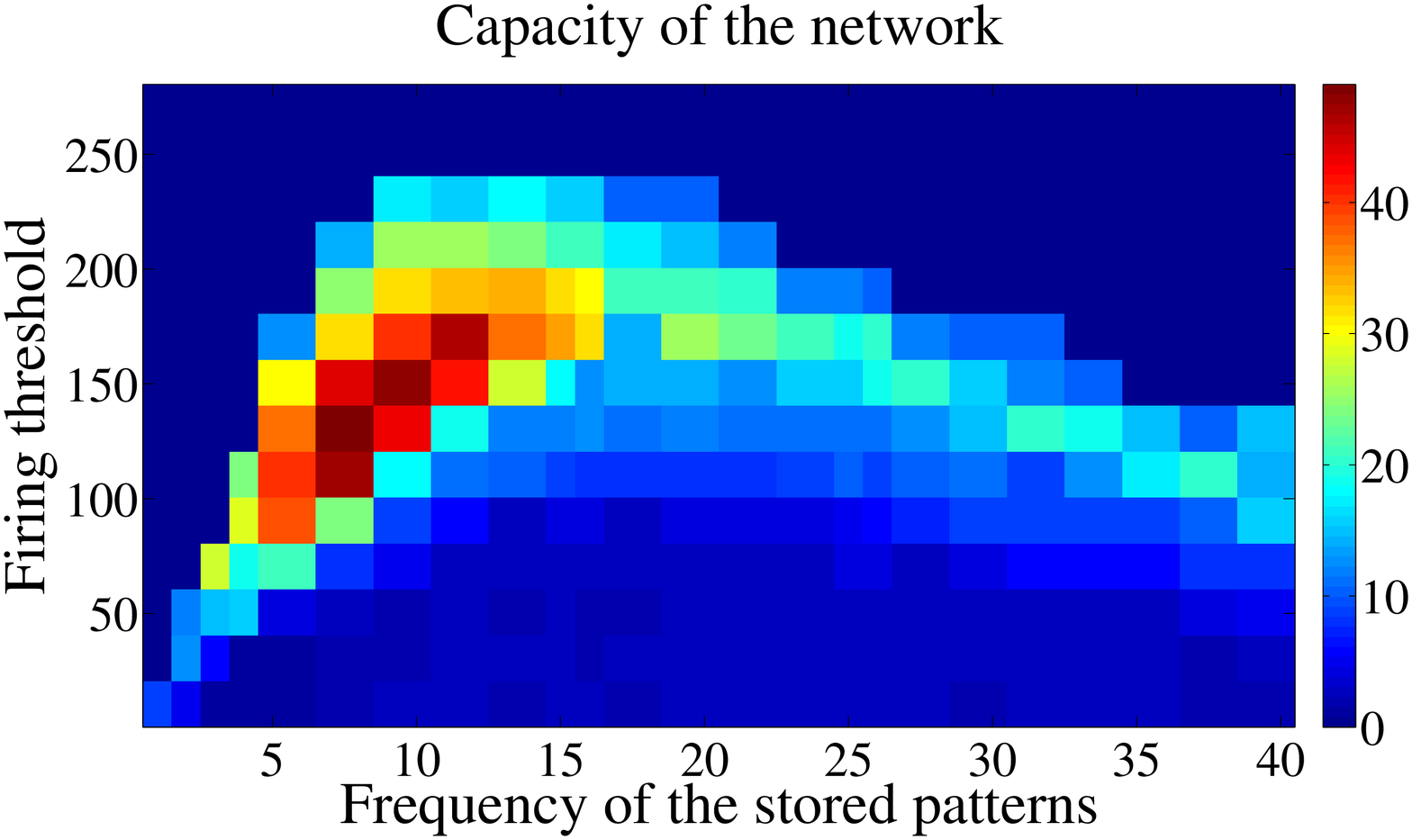}
b)
\includegraphics[width=4.5cm]{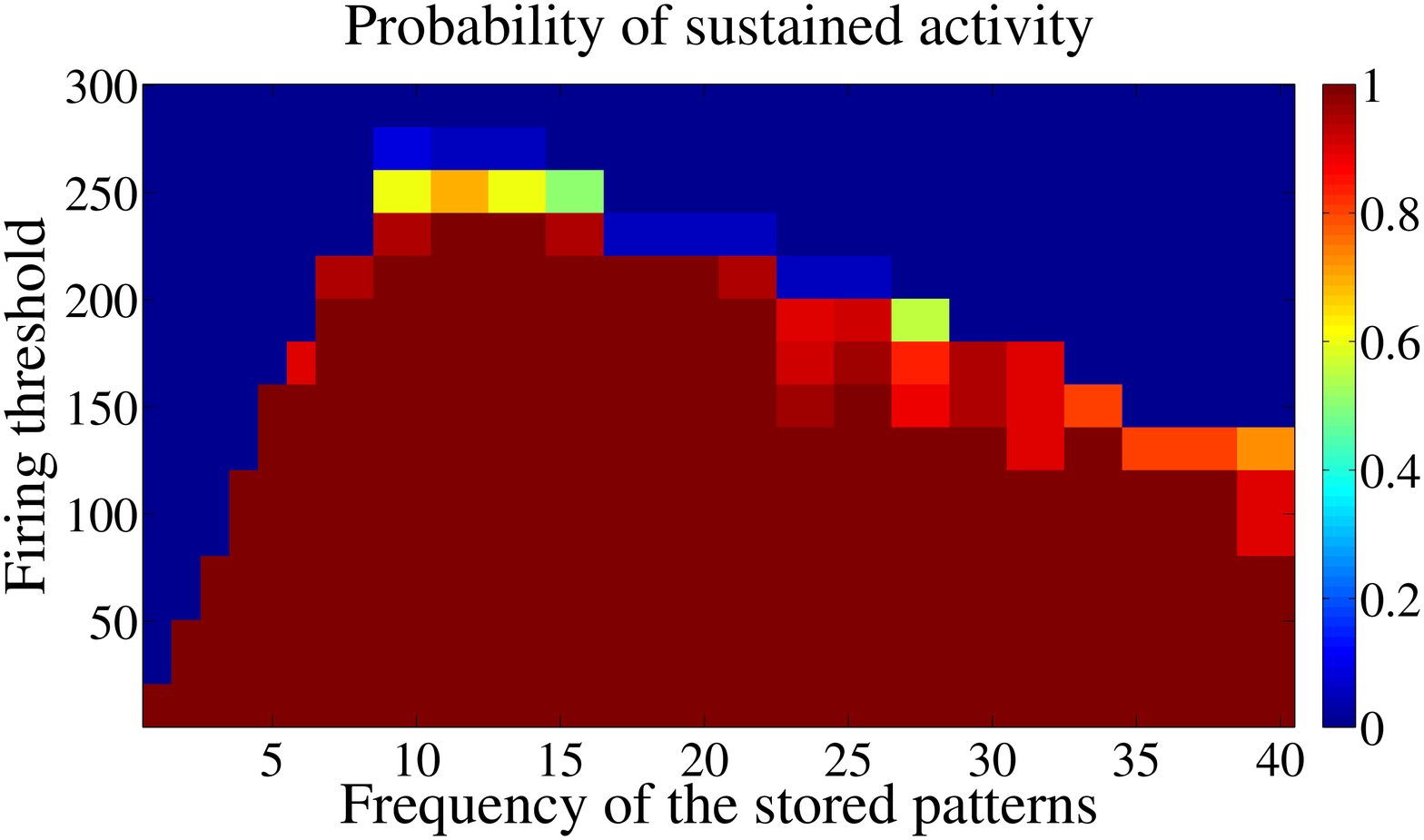}
\end{center}
\caption{a) Summary of the network capacity as a function
of the frequency and firing threshold. The  color legend shows the
number of correctly stored pattern.  b) Probability of self-sustained 
activity for a network with P=3 stored patterns in the plane $(\omega,\Theta)$. 
The color legend shows the region of the plane where a pattern
could be stored.}
\end{figure}
The Figure 4 shows results of numerical simulations averaged  over
50 runs, namely for different implementations of the network and
patterns to be stored.
The overlap $|m^\mu(t)|$ is reported in Fig. 4a  in a black and
white colored legend, along with the probability of self-sustained
activity, Fig. 4b as a function of $\omega_\mu$ at fixed firing
threshold ($\Theta=170$).
This gives an indication of whether or not the
initial stimulating spikes are sufficient to generate a persistent
spontaneous oscillatory activity regardless of the phases
alignment between neurons. In our analysis we did not considered
the non-persistent activity, that is the spontaneous dynamic
occurring in a short transient time, right after the initial
stimulating spikes. This means that black colored areas in Fig. 4
are not necessarily associated with an absence of spontaneous
activity, but only to an absence of long term activity. Hence, we
consider a successful pattern replay when the overlap, weighted
with the probability of long term sustained activity, is larger
than 0.5. This is reported in Fig. 4c, where we observe a large
interval of frequency with a good storage capacity.\\
We also investigate the role of the firing threshold $\Theta$.
Note that changing $\Theta$ in our model may correspond to a
change excitability. i.e a change of firing threshold or in global change in
the synaptic connections $J_{ij}$. Indeed, the
working range of the network depends on both the frequency of the
stored pattern, as well as on the threshold $\Theta$.
The capacity of the network is summarized in Fig. 5, where we report, in 
the plane frequency-threshold $(\omega,\Theta)$,
the number of perfectly replayed pattern (Fig. 5a) and the
probability of sustained activity (Fig. 5b).\\
Another important result is observed looking at neurons firing
activity. In Fig. 6 we see that by lowering the threshold
$\Theta$ below an optimal value, a burst of activity takes place
within each cycle, with phases aligned with the pattern. This open the possibility to
have a coding scheme in which the phases encode pattern's
information, and rate in each cycle represents the strength and
saliency of the retrieval or it may encode another variable. The
recall of the same phase-coded pattern with different number of
spikes per cycle accords well with recent observation of Huxter
{\em et al.} \cite{huxter_burgess} in hippocampal place cells,
showing occurrence of the same phases with different rates.  They
show that the phase of firing and firing rate are dissociable and
can represent two independent variables, e.g. the animals location
within the place field and its speed of movement through the
field.
The number of spikes per cycle as a function of the
threshold $\Theta$ is reported in Fig 7a, where a dependence
on the frequency of the replayed pattern is also observed in the
plane frequency-threshold.\\
\begin{figure}[ht]
\begin{center}
a)\;
\includegraphics[width=3.5cm]{1spikeprciclo.eps}
b)\;
\includegraphics[width=3.5cm]{2spikeprciclo.eps}
c)\;
\includegraphics[width=3.5cm]{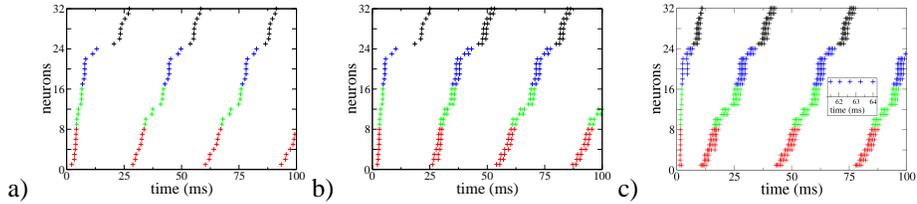}
\end{center}
\caption{%
Recall of the pattern $\mu=1$ for networks of 3000 neurons having
different values of parameter $\Theta$, namely
$\Theta=100$ in a), $\Theta=60$ in b) and $\Theta=20$
in c). Depending on the value of $\Theta$, the phase-coded
pattern is replayed with a different number of spikes per cycle. }
\label{fig_T}
\end{figure}
\begin{figure}[ht]
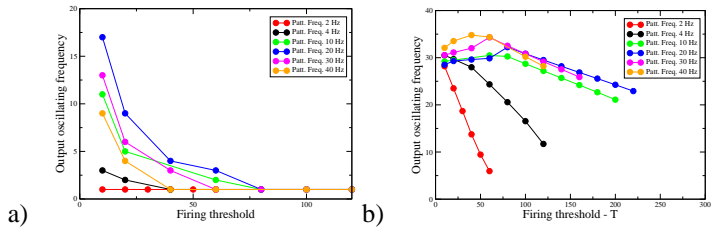

\begin{center}
a)\quad
\includegraphics[width=4cm]{spikeperciclo_last.eps}
b)\quad
\includegraphics[width=4cm]{freqout2_last.eps}
\caption{(a) The number of spikes per cycle as a function of the
firing threshold $\Theta$ for a network with one encoded
pattern, at different frequencies $\omega_\mu/2\pi$. (b) The
network oscillating frequencies as function of the firing
threshold. A stored  pattern is replayed at frequencies which
slowly decays with increasing $\Theta$ for most of the pattern
frequencies $\omega_\mu/2\pi$. 
}
\end{center}
\end{figure}
Lastly, in Fig. 7b, the dependence of  the  output frequency of collective
oscillations is studied as a function of 
$\Theta$. Notably, the stored phase-coded patterns are replayed in a compressed 
time scale for $\omega^{\mu}/2\pi< 30$ Hz. 
\section{Conclusions}
\label{sec_summary} We studied the storage and replay properties
of a network of spiking integrate and fire neurons, whose learning
mechanism is based on the Spike-Timing Dependent Plasticity. The
encoded patterns are periodic spike sequences, whose features are
encoded in the relative phase shifts between neurons. The proposed
associative memory approach, that replay the stored sequence, can
be a method for recognize an item, by activating the same
memorized pattern in response of a similar input, or could be a
method to transfer the memorized item to another area of the brain
(such as for memory consolidation during sleep). We systematically
quantify and compare the retrieval capacity of the network by
changing two parameters of the model: the frequency of the input
(encoding) patterns, $\omega/2\pi$, and the neuronal firing
threshold, $\Theta$. The response of the network changes by changing 
those parameters which, however, are not the only ones governing the 
spiking activity of neurons. Future works will consider a further analysis
of the model parameters and wheter to tune them to modify the network capability
in a controlled manner.

\end{document}